# Bell Inequalities And Hidden Variables Over All Possible Paths In A Quantum System

Warren Leffler
Department of Mathematics,
Los Medanos College,
2700 East Leland Road Pittsburg,
CA 94565
wleffler@losmedanos.edu

ABSTRACT

Bell's theorem rests on the following fundamental condition for a local system:

$$P(a,b \mid \alpha,\beta,\lambda) = P(a \mid \alpha,\lambda) P(b \mid \beta,\lambda).$$

Here $a$ and $b$ are the outcomes respectively for measurements $\alpha$ on one side, and $\beta$ on the other, of an experiment involving two entangled particles traveling in opposite directions from a source. The parameter $\lambda$ (the set of "hidden variables") represents a more complete description of the joint state of the two particles. Because of $\lambda$, the joint probability of detection is now dependent only on $\lambda$ and the local measurement setting of $\alpha$; similarly for the other side and the setting $\beta$. From this equation John Bell derived a simple inequality that is violated by the predictions of quantum mechanics, which is generally taken to imply that quantum mechanics is a nonlocal theory. But, by combining Richard Feynman's formulation of quantum mechanics with a model of particle interaction described by David Deutsch, we develop a system (the "space of all paths," SP) that (1) is immediately seen to replicate the predictions of quantum mechanics, (2) has a single outcome for each quantum event (unlike MWI on which it is partly based), and (3) contains the set $\lambda$ of hidden variables consisting of all possible paths from the source to the detectors on each side of the two-particle experiment. However, the set $\lambda$ is nonmeasurable, and therefore the above equation is meaningless in SP. Moreover, using another simple mathematical expression (based on the exponentiated-action over a path) as an alternative to the above equation, we show in a straightforward argument that SP is a local system. We show next that the famous GHZ argument fails in SP. Finally—building on a construct of Bernstein, Green, Horne, and Zeilinger (BGHZ)—we present an argument that there are just two mutually exclusive choices for quantum foundations: systems structurally similar to the space of all paths (such as MWI) or those that harbor action at a distance.

## I. INTRODUCTION

In his 1964 paper [1] John Bell developed a simple mathematical inequality that—along with similar inequalities by others, called "Bell inequalities"—eventually became a prominent feature of quantum foundations [2, 3, 4]. Bell inequalities follow from a condition that, in Bell's words [5], is assumed to incorporate "local causality" or "no action at a distance" into the analysis of correlations for two entangled particles. When experiments are carried out on spatially separated pairs of such particles, the results (as quantum mechanics predicts) will typically violate the numerical bound of a Bell inequality. This is said to imply that the observed correlations are "not locally explicable" (Ref. 5)—that they "cannot be explained by a locally causal theory, and are referred to as nonlocal correlations" [6]. As reported in one recent experiment, the two spatially separated systems are somehow tied together by influences that "propagate at least $10^7$ times faster than the speed of light" [7].

Interestingly, however, there is a tacitly assumed premise in Bell's simple mathematical argument, a premise that is easily recognized when it is pointed out. To see this (expanding on the point made in the above abstract), recall that the inequalities are based on the following notion of locality (Ref. 1, 5): Let $A$ and $B$ be two space-like separated systems involving two entangled particles that travel out in opposite directions from a source. Suppose that a measurement $\alpha$ is performed on system $A$, obtaining outcome $a$, and a measurement $\beta$ on system $B$, with outcome $b$. Let $P(a,b|\alpha,\beta)$ be the probability for the joint outcome. Now suppose further that $\lambda$ (the so-called "hidden-variables" parameter) represents a more complete description of the joint state of the two particles, one in which the joint probability factors into two independent local probabilities:

$$P(a,b|\alpha,\beta,\lambda) = P(a|\alpha,\lambda)P(b|\beta,\lambda). \tag{1}$$

Therefore, according to Eq. (1), the probability of detection at $A$ is now dependent only on $\lambda$ and the local measurement setting of $\alpha$; similarly for side $B$ and the setting $\beta$. This is sometimes called "the Bell factorizability condition" [8]. Bell explained the condition as a "decoupling" in terms of "causal" factors (Ref. 2). Ever since Bell's original argument, Eq. (1) has been taken as the fundamental condition for locality. As it is sometimes expressed, "locality requires that some set of data, $\lambda$—made available to both systems, say, by a common source—must fully account for the dependence [between the two sides]." [9]

On the basis of Eq. (1), Bell derived in a few short steps an inequality that is now taken to be a necessary condition for a local theory. Quantum mechanics violates this inequality. Therefore the conclusion is that quantum mechanics is necessarily a non-local theory: "no local theory can make the [quantum mechanical] predictions for the results of experiments carried out very far apart" [10]. We will refer to this as the "non-locality conclusion."

But suppose that we take the hidden variables to be the set $\lambda$ of all possible paths $x(t)$ from the source to the detectors on each side of a two-particle experiment. (As Bell stated in Ref. 1: "It is a matter of indifference in the following whether $\lambda$ denotes a single variable or a set, or even a set of functions, and whether the variables are discrete or continuous.")  This set, however, is nonmeasurable—that is, there is no countably-additive and translation-invariant measure on the underlying space  [11, Appendix B]. Moreover, the subsequent steps in the derivation of Bell's celebrated theorem (which involve a translation-invariant, countably-additive probability distribution, $\rho(\lambda)$) are meaningless in a space that has no translation-invariant and countably-additive measure.

Therefore, there is only one way to salvage (if anyone is so inclined!) the above non-locality conclusion of Bell's theorem. That is, for Bell's non-locality conclusion to hold for quantum mechanics one must claim that quantum mechanics requires a measurable space. But does it, really? We shall show that this claim is false, by developing a straightforward synthesis (the "space of all paths," SP in Appendix A) of Richard Feynman's path-integral formulation of quantum mechanics with a model of particle interaction described originally by David Deutsch to eliminate single-particle interference paradoxes [12]. The underlying space in SP is nonmeasurable, and yet the system replicates the predictions of quantum mechanics (and therefore violates Bell's inequality). Moreover, as we show in Sec. II, the space of all paths operates throughout by local causality.

Although the space of all paths is nonmeasurable, the sum over all possible paths of the exponentiated-action terms, $\exp(iS[x(t)]/\hbar)$ (where $S$ is the action over a path $x(t)$ in configuration or spin space), leads to probability outcomes equivalent to those of standard quantum mechanics [13,14]. (By standard quantum mechanics, SQM, we mean a mathematical representation of quantum phenomena generally based on some form or other of Lebesgue square-integrable wavefunctions over a separable, Cauchy-complete Hilbert space.) The set $\lambda$ of paths in SP from the source to the detectors is available to both sides and (as we show in Sec. II) accounts for the dependence between the two sides. Of course $\lambda$ depends on the experimental setup but the set of possible paths is fixed throughout all runs of the experiment using that setup. What varies randomly at each run of the experiment are the two paths traveled by the two entangled "tangible" particles (point (2) below).

Note that *the essential requirement of a physical theory is certainly not that the underlying space must have a translation-invariant, countably-additive measure*. Instead, what is required is that we must be able to compute probabilities of quantum events. Now, there was a time when the path integral (the sum over all paths) was felt to lack mathematical rigor, but that has greatly changed in the seventy years since Feynman discovered it. There are presently many rigorous approaches to the path integral, and its validity has been shown over a wide class of configurations and potentials, for both configuration space and spin space, with no contrary results [15]. (This includes a closed-form path-integral implementation of the experimental setup—a standard Rarity-Tapster two-particle interferometer—that we analyze in Sec. II.) Moreover, the path integral has become nearly indispensable in quantum field theory. Also, one should keep in mind that Bell's transparent argument is clearly valid in a measurable space, with the consequence that correlations over widely separated regions then occur in a way that is utterly mysterious. As John Wheeler and Martin Gardner put it, the particles would have to "remain connected, even though light years apart, by a nonlocal sub-quantum level that no one understands ...." [16]. On the other hand, *one might claim that one of the fundamental purposes of a scientific theory is to explain correlations*. Indeed, as Bell noted (Ref. 5), "the scientific attitude is that correlations cry out for explanation."

The SP approach explains the correlations in a straightforward way (Sec. II) in terms of all possible paths being traveled within the system.

SP is a theoretical framework that has the following properties:

1. It replicates the predictions of standard quantum mechanics, as already mentioned.

2. In SP each "tangible" (ordinary) particle traveling between two points in a space (whether in configuration or spin space) generates a "shadow stream" of Deutsch's counterpart particles traveling all possible paths between the points. The tangible particle randomly takes one of the paths, and the shadow particles the others (hence the tangible belongs to its shadow stream). Tangible particles only interact with shadow counterparts of the same type—tangible photons with shadow photons, tangible electrons with shadow electrons, etc.

3. The probability of an event in SP joining two points $x_1$ and $x_2$ in a space is proportional to the square of the length of the vector sum over all unit vectors corresponding to paths from $x_1$ to $x_2$:

$$\sum_{\substack{\text{over all paths} \\ \text{from } x_1 \text{ to } x_x}} \exp(iS[x(t)]/\hbar),$$

where $S$ is the action over a path $x(t)$ from $x_1$ to $x_2$—whether in configuration or spin space (that is, it is just complex conjugation of the vector sum). Although this probability function is not translation-invariant and countably-additive in SP, and so cannot be employed in carrying out Bell's argument, it nevertheless yields outcomes equivalent to those of standard quantum mechanics. Following Feynman [17], we can graphically represent the vector sum over the shadow stream (for the event given by a photon traveling from $x_1$ to a mirror and then to $x_2$) by the "long" arrow as in Fig. 1 below. (But of course Feynman did not describe this construct in terms of a shadow stream.)

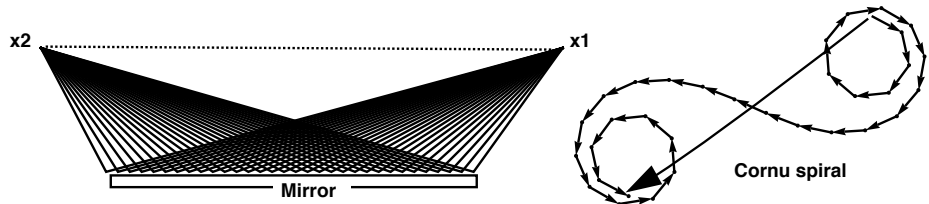

Fig. 1 Here we imagine a tangible particle, say a photon, accompanied by a stream of shadow photons. Each particle in the stream travels from x1 to a mirror, where it is then reflected up to point x2. The path of each such particle is associated with the exponentiated action, exp(iS[path])/ h-bar), over its path. This can be viewed as a "clockpointer" whose hand rotates at the classical frequency of the light. The rotation starts when the photon is emitted, and stops when the photon arrives at x2. For each particle when it reaches x2, the final position of its clockpointer produces a unit vector for that path. When the unit vectors are added head-to-tail they produce a vector sum, the "long" arrow in the Cornu spiral. The clockpointers associated with particles traveling paths of roughly equal time end up pointing in roughly the same direction, adding up constructively in the vector sum, while those in the "curled-up" portions tend to cancel.

4. In SP there is a single outcome for each quantum event. This fact makes it an appropriate system for examining Bell-type arguments for two or more entangled particles (unlike MWI [18,19], on which Deutsch based his model of particle interaction).

5. The above set $\lambda$ of all possible paths is a complete state in the EPR sense [20], richer in content than the standard quantum state, and it provides a commonsense, local explanation of the correlations predicted by quantum mechanics. We justify this in detail in Sec. II.

6. There is an infinitely-many-to-one morphism from SP onto the standard Hilbert-space representation of quantum mechanics, in which homotopy classes of paths are mapped onto basis vectors in the Hilbert space (an example is given in Sec. II). In other words, the standard representation is "incomplete" in the sense of EPR. (Recall that two paths with the same endpoints are homotopic when they can be continuously transformed to each other.) The morphism collapses all relevant information about quantum interactions in SP onto the image space (the vector space of standard quantum mechanics), in much the same way that a modular homomorphism collapses information about the integers onto the integers mod *n* (though the vector-space image is not finite, unlike the space of integers mod *n*).

7. Bell type arguments (such as GHZ [21]) for more than two entangled particles cannot be carried out in SP. This is important because, unlike Bell's theorem, the more-than-two-particle arguments are not statistical in nature. Although a detailed justification is outside the scope of this paper, here is a quick sketch of why GHZ fails in the space of all paths. It hinges on the notion of an "observable."

Consider, for example, spin-½, which is represented in SQM by a two-dimensional Hilbert space. In SP, on the other hand, one views a spin-½ particle (such as an electron) in terms of local realism as (say) a tiny spinning dipole magnet in Fig. 2 below. Any mathematical representation of spin-½, such as SP or SQM, must account for several kinds of fundamental experiments, which involve sending a particle through a sequence of Stern-Gerlach (SG) devices oriented in various directions. [22] As each particle in the stream travels through an SG magnetic field it enters randomly one of the two homotopy classes of path in $SO(3)$. When it exits the field the interference effects produce a landing in one of the two "intrinsic spin" positions of "up" or "down" (the observable) on the detecting plate. If the tangible enters a second SG-field oriented in the same direction it continues in the same homotopy class as it was in when it exited the first SG-field; otherwise it randomly enters one of the two homotopy classes, and so forth. These results are quantified by the sum of the exponentiated action, $\exp(iS[path]/\hbar)$, over each possible spin path in $\mathbb{R}^3 \times SO(3)$—for spin the action is the sum of a magnetic field action and a "top" action. It is easy to see that this SP representation replicates the outcomes for the above fundamental experiments.

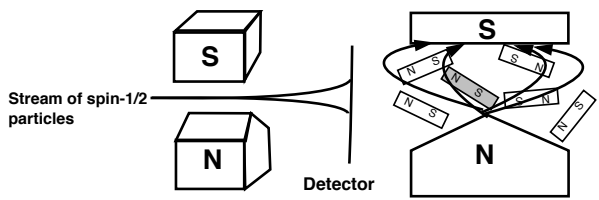

**Fig. 2** The darker particle is a tangible particle traveling in a shadow homotopy class with other spin 1/2 particles (tiny dipole magnets)

SQM and SP have the same observables, but in SP the element of reality of the tangible particle is not the observable, which is the result of interference with infinitely many other particles in the stream; rather it is the actual top-like spin of the tangible particle (the darker colored one in the figure) as it exits the SG field, which (because of interference effects) we can never detect directly. The observable merely "points to" the element of reality, as it resides in a swirl of interference effects. The tangible's path is what is associated at each instant with the tangible's classical property, its "SP element of reality." But the GHZ argument identifies EPR's element of reality with the notion of an observable, and shows this leads to a contradiction. Yet, EPR never said an element of reality was identical to an observable (they gave a sufficient condition based on 100 percent predictability—which is consistent with how we just described the concept as a pointer in SP).

Thus the GHZ argument fails in SP because path configurations do not form a number field in which one can carry out arithmetic operations such as multiplication,

etc.—which is how GHZ was able to predict the outcome of various intrinsic spin outcomes given the results of others. Incidentally, in certain respects the notion of an observable in SP resembles Bell's idea of a "beable," where he writes: "The beables of the theory are those elements which might correspond to elements of reality, to things which exist. Their existence does not depend on 'observation'."[23]

It seems surprising that no one until now has noticed the hidden premise of measurability in Bell's definition of locality (in contrast to a critic of Bell's theorem who might claim, for instance, that the probability distribution somehow contains inappropriate parameters or, alternatively, omits parameters—which are mathematically vague assertions that invariably lead to mistakes buried deep within the formalism [24]). But perhaps this is at least partly because of two natural assumptions: First, the space in which we conduct quantum experiments is assumed to be Euclidean space, which is, of course, measurable—although, as we show (Sec. II), the analysis of the results for such experiments requires a deeper reality. Second, in studying quantum mechanics virtually everyone is trained initially in some form of a Hilbert-space approach to the foundations, which assumes a space of square-integrable wavefunctions or kets. Thus in introductory quantum mechanics, square-integrable solutions to Schrödinger's equation are what are typically taken to be physically realizable states [25]. Measurability is of course presupposed in such approaches to the foundations.

There are, surprisingly, essentially just two, mutually exclusive, choices for quantum foundations: systems structurally similar to the space of all paths or systems that harbor action at a distance. One can see this by building on an important result contained in a paper of Bernstein, Green, Horne, and Zeilinger (BGHZ), "Bell theorem without inequalities for two spinless particles" [26]. For their proof to go through, BGHZ found it necessary to augment it with a premise called "emptiness of paths not taken," EPNT. Of course SP is, in a sense, the denial of EPNT. By extending the EPNT hypothesis to include homotopy classes of paths, one has, mutatis mutandis, BGHZ's action at a distance. Although one could deny EPNT and still have a nonlocal theory (for example, David Bohm's pilot-wave theory, which in effect "fills" the paths with a "pilot wave") [27], we nevertheless have the result that "SP" implies "no action at a distance," and "not SP" implies "action at a distance."

Finally, although it is beyond the scope of the present paper to develop the idea further, we also note that that the local system of SP, despite its immunity to Bell-type inequalities, continues to enable the quantum advantage in such areas as communication complexity problems, device independent quantum cryptography, and so forth. It does this by virtue of a stream of particles traveling infinitely many paths in a quantum space, each particle following classical dynamics along its respective path and contributing to the outcome when observed (the observable being, as noted, a sum of exponentiated action terms—which is equivalent to the SQM observable).

## II. A COUNTEREXAMPLE TO BELL'S THEOREM BASED ON THE HIDDEN REALITY OF SHADOW PARTICLES

As noted in Sec. I, SP is a system that replicates the predictions of quantum mechanics (and so violates Bell's inequality), but one in which Bell's proof breaks down. We will now show that, furthermore, SP is a local theory.

As is clear from Sec. I, our argument that SP is a local system will not employ the Bell factorizability condition of Eq. (1), since this equation only makes sense in a measurable space. There is, however, another simple mathematical condition for locality that we can use—one that is every bit as representative of locality as the factorizability condition. This condition is based on the fact that we can associate a "clock-pointer," $\exp(iS[path]/\hbar)$, with each particle in a shadow stream as the particle travels its path from the source to a detector (as in Fig. 1, above).

To help explain this condition, we will now describe a gedanken, toy system in which two classical objects (billiard balls, tennis balls, bullets, etc.) proceed out from a source, traveling in opposite directions to forks located at equal distances along each path, the fork on each side leading to upper and lower branches. This is a local system, in which we imagine that each object carries a "clock-pointer," $\exp(iS[path])$, as the object on each side travels its path out from the source. The two pointers are set randomly at the source event, but point initially in the same direction. We also suppose that the pointers rotate counterclockwise at identical rates (given by the action—notice that we do not need the factor of $1/\hbar$ in this classical example) as the objects travel along their respective paths (just one path on each side in this classical setup). Just before the objects reach a fork point, "measurements" are carried out—$\alpha$ on the left side, $\beta$ on the right, where $\alpha$ and $\beta$ are one of the angles 0, $2\pi/3$, $4\pi/3$. For example, a setting of $\alpha$ ($0 < \alpha < 2\pi$) slightly lengthens the path of the particle on the left so that its clock-pointer further rotates by the additional amount $\alpha$ (the angle $\alpha$) over that stretch of the path; and similarly for $\beta$ on the other side. When the objects reach a fork, identical bivalent functions send them along the upper or lower branches at the fork, depending on whether their clock-pointers are pointing between 0 and $\pi$, or between $\pi$ and $2\pi$.

Because the action is the same over congruent paths (a point that holds also in the quantum mechanical case below), the correlation of position between the sides—the objects ending up on corresponding forks: (upper, upper) or (lower, lower)—is 100% when $\alpha = \beta$. When $\alpha \neq \beta$, the probability of correlation is 1/3, as shown in Appendix C. Here we note that for, say, $\alpha > \beta$ when the object arrives at a fork, the pointer on the right side is directed at some angle $\gamma$, while on the left side it is at angle $\gamma + (\alpha - \beta)$. The toy system obviously operates in a measurable space. Thus the correlations, being locally explicable, must satisfy Bell's inequality, but the toy system illustrates why the outcome on, say, the left side in a local theory can involve the measurement $\beta$ on the right side (which might at first glance seem contrary to locality). This is because the outcome is accomplished by congruent paths (apart from the measurements) and by clock-pointers—built-in "genetic" instructions. As we show below, this condition transfers also to the quantum setup, which has infinitely many possible paths. Because of the clock-pointers, there is no need for information to be sent across the origin in order to coordinate the correlations. (Interestingly, the value 1/3 may be a lower bound for a classical system incorporating the above measurements: It is considerably less, for instance, than the value of 1/2 for Mermin's famous red-green version of Bell's argument [28], which employs similar measurement settings.)

Now it turns out that the SP account of the correlations in a two-particle interferometer almost exactly parallels that for the toy system. The crucial difference is that in SP the infinitely many clock-pointers involved—exponentiated-action terms, $\exp(iS[x(t)]/\hbar)$—are unit vectors in the complex plane that we can combine head-to-tail (as in Fig. 4 below). We add them in this way because of interference effects, whereas there are of course no such effects in the toy system. The associated probability is then a function of the wave intensity of the streams on each side involving the interference effects (the sum over infinitely many paths).

Throughout the following discussion we will cite postulates for SP (described in Appendix A) that are intuitive and clearly consistent with experiment. The experimental device that we analyze using these postulates is a Rarity-Tapster two-particle interferometer (Fig. 3), which is especially suited to our argument. But, although it is beyond the scope of the present paper to discuss this issue in detail, it is the case (so we claim) that a similar analysis applies to any experimental arrangement involving a pair of entangled particles (at the end of this section we quickly sketch a similar argument for spin ½).  *In any case, we emphasize that it takes only one counterexample to overturn a general theorem such as Bell's.*

We of course know in advance that, since SP is equivalent to SQM, both systems yield the same probability for the correlations in our interferometer. But in SQM the mathematical implementation of the Rarity-Tapster interferometer is based on a finite-dimensional Hilbert space of square-integrable wavefunctions—a measurable space. The conventional picture of the interferometer (on the left side of Fig. 3) seems to suggest just finitely many paths ([29]), unlike the infinitely many possible paths indicated in the SP picture on the right side (though of course there are infinitely many kets in the associated vector space). But it is important to realize that the conventional picture is not what actually justifies the SQM representation. Rather it is that an SQM finite-dimensional Hilbert-space implementation accounts for the experimental outcomes. It has a serious drawback, however: It is incapable of explaining how the results follow from an inner mechanism [30]. Perhaps much worse, Bell's theorem is valid in the SQM representation, with the result that the predicted correlations are completely inexplicable in intuitively appealing terms (as noted in the Wheeler-Gardner quote in Sec. I). On the other hand, in the SP representation, based on Feynman's path-integral formulation, we can also coherently and simply replicate the predicted outcomes, but in a way that is both intuitive and locally explicable.

Thus consider the Rarity-Tapster interferometer of Fig. 3. On the left is the conventional diagram, and on the right is how it is pictured in SP, where the possible paths are suggested when a source event generates a pair of entangled tangible particles.

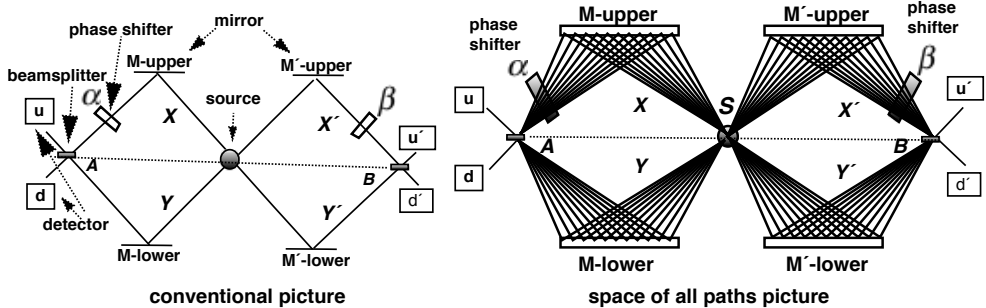

**Fig. 3** At each run of the experiment, paired spinless particles travel first to floor- and ceiling-mirrors (M-M´ lower and M-M´ upper) on each side, from which they are then reflected to 50-50 beamsplitters at A and B.

For any two-particle interferometer, interference can only occur if there is a certain amount of positional uncertainty in the source, unlike with single-particle interference, which requires a point source [31]. The Rarity-Tapster device handles this as follows: At each run of the experiment, paired, spinless particles travel first to floor- and ceiling- mirrors (M-M´ lower and M-M´upper in Fig. 3), from which they are then, on the respective sides, reflected to 50-50 beamsplitters at *A* and *B*. The setup generates infinitely many possible paths, forming a nonmeasurable space. Some of the possible paths are pictured in Fig. 3.

As pointed out earlier, because of the equivalence between SP and SQM, both systems yield the same probability for the correlations. In the SQM vector-space approach, however, one usually begins by defining the quantum state of the pair of entangled particles as, say,

$$|\psi\rangle = (|X\rangle_L |Y'\rangle_R + |Y\rangle_L |X'\rangle_R)/\sqrt{2},$$

where the ket $|X\rangle_L$ denotes the particle on the left traveling along path X, etc. But in SP it is convenient, from the outset, to work with amplitudes for joint detection.

We let $\lambda = L \cup R$ be the set of infinitely many possible paths going from the source to the two separated systems represented by A and B at the beamsplitters in Fig. 1, with $L = X \cup Y$ and $R = X' \cup Y'$. Then (as noted earlier) $\lambda$ is the set of "hidden variables" comprising the possible paths $x(t)$ from the source to the beamsplitters.

An expression such as $\langle A|X\rangle_L$ below denotes the amplitude for a stream of particles traveling along paths in the upper left stream from the source S to the beamsplitter at A (here the measurement setting $\alpha$ is left implicit). The amplitude $\langle A|X\rangle_L$ is proportional to the sum

$$\sum_{\substack{\text{over all upper paths} \\ x(t) \text{ from } S \text{ to } A}} \exp(iS[x(t)]/\hbar).$$

Similarly for the other amplitudes. We will now illustrate in SP one possible approach to computing the probability for the correlations between the sides—the outcome being equivalent to that of SQM.

In computing the amplitude for the entangled particles to flash the same at the detectors—both "up" or both "down"—one factors in $i$ (= $\exp(i\cdot\pi/2)$) for reflection from a beamsplitter. Also, in SQM one assumes that, because of conservation of momentum, the two tangibles take oppositely directed paths. This is the case in SP too, although we will discuss in a moment another rationale for this in SP. But in both SQM and SP we have, both "up" detectors flash: $\quad i \cdot \langle A|X\rangle_L \cdot \langle B|Y'\rangle_R + \langle B|X'\rangle_R \cdot i \cdot \langle A|Y\rangle_L,$ (2)

<small>left top path to up detector   right bottom path to up detector   right top path to up detector   left bottom path to up detector</small>

both "down" detectors flash: $\quad \langle A|X\rangle_L \cdot \langle B|Y'\rangle_R \cdot i + \langle B|X'\rangle_R \cdot \langle A|Y\rangle_L \cdot i.$ (3)

<small>left top path to down detector   right bottom path to down detector   right top path to down detector   left bottom path to down detector</small>

Using the congruence of the paths (apart from the phase shifts, $\alpha$ and $\beta$) we can represent these amplitudes (which are just complex numbers) by

$$\langle A|X\rangle_L = re^{i\theta}e^{i\alpha}, \langle A|Y\rangle_L = re^{i\theta}, \langle B|Y'\rangle_R = re^{i\theta}, \langle B|X'\rangle_R = re^{i\theta}e^{i\beta}. \quad (4)$$

Thus the amplitude that both detectors flash the same is

$$ire^{i\theta}e^{i\alpha} \cdot re^{i\theta} + re^{i\theta}e^{i\beta}i \cdot re^{i\theta} = ir^2 e^{i2\theta}(e^{i\alpha} + e^{i\beta}). \quad (5)$$

When we take the absolute square (multiply by the complex conjugate), we have

$$P(\text{both-flash-same over } \lambda | \alpha, \beta) \propto r^4(e^{i\alpha} + e^{i\beta})(e^{-i\alpha} + e^{-i\beta}) = r^4(2 + e^{-i(\alpha-\beta)} + e^{i(\alpha-\beta)}). \quad (6)$$

Performing simple algebraic operations over the complex numbers, and choosing a suitable proportionality factor, we see that the probability for the detectors to signal the same, both "up" or both "down," is thus

$$\cos^2 \frac{(\alpha - \beta)}{2}. \tag{7}$$

Now expressions such (2) or (3) seem to suggest that the tangible particles must somehow communicate across the source in order to coordinate the predicted correlations. But of course by itself such an expression can show no such thing. A conclusion of that kind requires Bell's argument, which breaks down in SP.

Because SP is a nonmeasurable space, our argument that SP is a local system cannot be based on the Bell factorizability condition, but (as noted earlier) there is another simple mathematical approach based on "clock-pointers," exponentiated-action terms. This approach parallels the one above for the toy system involving classical objects.

First, by postulate 5, we stipulate that when two entangled tangible particles are generated by a source event in a two-particle interferometer, their clock-pointers are initially aligned in the same direction. We also extend this condition (in effect an "initial action-instruction") to all counterpart "twin" shadow particles in their respective shadow streams on each side. Postulate 5 is clearly consistent with experiment.

In a two-particle interferometer we have four streams (homotopy classes of paths in SP), two on each side (an upper and a lower shadow stream). As we show in a moment, when the streams reach a beamsplitter, where the tangibles go next (to which detector) is a consequence of local, independent action on each side, the outcome governed by the sum of exponentiated-action terms in the complex plane (the "long" vector in Fig. 4 below).

Now in any two-particle interferometer similar to ours, the emission angle subtended at the source by the upper paths must be sufficiently acute for correlations to occur [32]. Thus, if the angle is too large, then each side corresponds to single-particle interference, and the correlations disappear. Also, in SP a tangible particle always pursues a single (randomly taken) trajectory by Postulate 1. This means that there are just two choices for what may be termed the "upper" tangible particle: It can only take a path in the homotopy class $X$ or in the class $X'$ in the SP path-setup of Fig. 3—which path it does take varies randomly by Postulate 1 at each trial of the experiment. If the upper tangible takes, say, a path in $X$ then the other, lower tangible must by conservation of momentum take a linked, oppositely directed path in $Y'$. Similarly, if the upper takes a path in $X'$, then the other tangible is linked to take a path in $Y$. This implies that the particles taking paths in the homotopy classes $X$ and $X'$ are from different streams; likewise for those taking paths in $Y$ and $Y'$. In other words, $X$ and $Y$ are generated by different tangibles (in our set of two entangled tangible particles); likewise so are $X'$ and $Y'$. By postulate 3, the composite amplitude of the shadow streams generated by different tangibles is the product of the separate amplitudes.

In Fig. 4 (the left-half of the SP picture in Fig. 3) a tangible particle travels from a source to a mirror (above or below) and is then reflected to a beamsplitter at $A$. As we know, when the particle encounters a 50-50 beamsplitter, half the time on average it is transmitted, and half of the time reflected. In SQM, there is no explanation for this quantum observable, just a description of its behavior [33, 34]. As Feynman once wrote from the perspective of SQM (though he was in fact describing a path-integral account), "I am not going to explain how the photons actually 'decide' whether to bounce back or go through; that is not known. (Probably the question has no meaning.)" [35] In SP, however, we can explain the result quite simply in terms of "machinery" behind the observable. What's ironical is that our explanation draws on Feynman's path-integral machinery. For convenience, we will confine our discussion to $\mathbb{R}^2$.

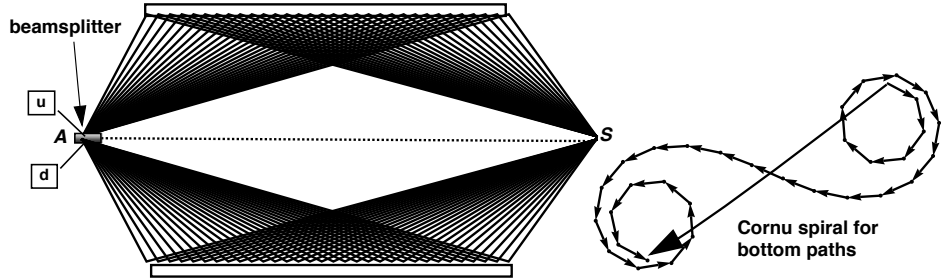

**Fig. 4** A free tangible particle in a two-particle interferometer travels from a source to a beamsplitter, accompanied by shadow counterparts. The particles travel to a mirror (upper or lower) and are reflected back to the beamsplitter at A. The tangible randomly takes one of the paths in either the upper or lower homotopy classes. This generates over a finite set of paths two Cornu spirals (only the one for the bottom paths is shown) consisting of unit vectors, exp(iS[path]/h-bar), added head-to-tail. The sum of the unit vectors is represented by the "long" vector.

In our explanation, as the tangible particle in a two-particle interferometer travels from the source to the beamsplitter at $A$, it is accompanied by two shadow streams of particles. The observable (that is, whether the tangible is transmitted or reflected) results from interference effects within and between the streams. The amplitude for this event is a product of two sums, an upper and a lower one. The lower sum is represented (proportionally) by the "long" vector in the Cornu spiral of Fig. 4 (which depicts a finite subset of the sum of unit vectors—exponentiated-action terms—over the infinitely many possible paths) [36]. The product of both sums (the upper and lower "long" vectors) is a complex number of the form, $re^{i\theta}$.

In postulate 4 we posit that there is a rotation $\Gamma$ of the unit circle in $\mathbb{R}^2$ associated with each beamsplitter (an unknown variable, of course). If the angle $\theta$ in $re^{i\theta}$ above is in the interval $\Gamma(\{(\cos\phi,\sin\phi):0\leq\phi\leq\pi\})$, then the tangible is transmitted by the beamsplitter, otherwise it is reflected. In this way we obtain the observable associated with a 50-50 beamsplitter. This outcome occurs by purely local interaction. Postulate 4 is obviously consistent with experiment, and it explains the experimental results when a single tangible particle is sent to a beamsplitter in a two-particle interferometer (in a single-particle interferometer—also discussed in postulate 4—a similar process occurs, but in that case there is just one counterpart shadow particle also traveling to the beamsplitter on a separate path, not infinitely many as in the two-particle case).

Also, each path in the stream in the lower half on the left side of the SP picture in Fig. 3 corresponds in a one-to-one fashion to a congruent path in the lower half of the right side. Congruent paths have equivalent exponentiated actions, so the sum on the left side over all the lower paths will be the same as on the right. That is, $\langle A|Y\rangle_L = \langle B|Y'\rangle_R$. Therefore (substituting "equals for equals" into expression (2) or (3) and applying postulate 3), the amplitude for both detectors to signal the same (that is, both "up" or both "down") is,

$$i\left(\langle A|X\rangle_L\langle B|Y'\rangle_R + \langle B|X'\rangle_R\langle A|Y\rangle_L\right) = i\left(\langle A|X\rangle_L\langle A|Y\rangle_L + \langle B|X'\rangle_R\langle B|Y'\rangle_R\right). \qquad (8)$$

A product such as $\langle A|X\rangle_L \langle A|Y\rangle_L$ on the right half of Eq. (8) is of the form $re^{i\theta}$, for some $\theta$. By postulate 4 in SP the local interference effects associated with such an amplitude at the beamsplitter are what determine where the stream containing the tangible particle goes next after leaving the beamsplitter—that is, "up" or "down" in our setup. This is an independent physical process on each side, operating the same in the multi-particle case as in the single-particle one. (Indeed it closely resembles the similar situation in the above toy example.)

In the two-particle interferometer of Fig. 3, the interference on a side comes from a product of two local amplitudes—such as $\langle A|X\rangle_L \langle A|Y\rangle_L$ on the left side of the SP picture in Fig. 3—a composite amplitude, but of the form $re^{i\theta}$. This composite amplitude is independent of the similar one $\langle B|X'\rangle_R \langle B|Y'\rangle_R$ on the other side. That is the content of postulate 4. The sum

$$\langle A|X\rangle_L \langle A|Y\rangle_L + \langle B|X'\rangle_R \langle B|Y'\rangle_R \qquad (9)$$

is therefore the analogue in SP of the Bell factorizability condition.

Thus, although Bell's argument breaks down in SP, we see that an intuitively appealing "local condition" (expression (9)) nevertheless holds in the space of all paths, while at the same time the predicted correlations violate Bell's famous inequality. The dynamics are similar to what takes place in the above toy, classical system, in which we imagined that a pair of "twin" classical objects travel in opposite directions from a source along a single path on each side. The tighter quantum mechanical results (that is, a smaller probability) stem, however, from the fact that the sum (and then product) over exponentiated-action terms issuing from interference effects leads to a complex number (a plane wave) such as $re^{i\theta}$ (whereas there are no interference effects in the classical, single-path system).

To recapitulate: When entangled particles travel out from a source event along a path, their "clock-pointers," $\exp(iS[path]/\hbar)$, are set pointing in the same direction (postulate 5). This is the initial instruction. As the upper and lower streams of particles (tangible and shadow) on each side travel to a beamsplitter their clockpointers (governed by the action-instructions) rotate at the same rate over congruent paths. These represent built-in, "genetic" instructions for each side. When the two streams on a side reach a beamsplitter, the product of the upper and lower sums over the interference effects (proportional to the long arrow in a Cornu spiral such as that of Fig. 4) determines (by postulate 4) where the tangible particle will go next, up or down. This closely parallels the above toy system and is a purely local process both for a single or for a multiparticle event.

The particles in a stream follow blindly their initial programming—their "genetic action-instructions," together with the path-dynamics governed by the laws of physics pertaining to the action over a path. The behavior at a beamsplitter in a two-particle interferometer is thus independent of what occurs on the other side. No information needs to be sent across the source. In the two-particle case, the effects are governed by a local sum of paths in the product of two amplitudes such as, say, $\langle A|X\rangle_L \langle A|Y\rangle_L$ on the left side. In other words, the outcome on each side is the result of local interaction involving exponentiated-action terms over all paths on a side, and is independent of what happens on the other side. In the Rarity-Tapster interferometer, the correlations between the sides are the consequence of various correspondingly congruent paths, leading to correspondingly equivalent exponentiated-action terms, apart from the phase-shifts. Each phase shift affects just its side independently, as in the toy system, and yet the sum of independent amplitudes (a product of sums over upper and lower paths) on each side leads (when the complex conjugate is taken) to the joint probability of the correlations. **QED**

Deutsch showed in the context of MWI how shadow particles eliminate single-particle interference paradoxes. We've now extended a similar construction to show the same for entangled particles. (There is possibly an infinitely-many-to-one morphism from SP to MWI—because paths are continuous in SP—but this topic is beyond the scope of the present paper.)

As Bell told Jeremy Bernstein [37], the predicted correlations seem to demand something like the "genetic" hypothesis—where identical twins carry with them identical genes. "This is so rational that I think that when Einstein saw that, and the others refused to see it, *he* was the rational man … The other people, although history has justified them, were burying their heads in the sand … So for me, it is a pity that Einstein's idea doesn't work. The reasonable thing just doesn't work."

On the contrary, as we have just seen, the reasonable thing does work in SP. Quantum mechanics is thus not quite as weird as it might seem if it allowed, across arbitrary distances, instantaneous action that takes place according to mysterious dynamics that no one understands. It does, however, rest on a hidden reality of shadow particles. Although the particles in a stream go every which way, they nonetheless obey classical action dynamics along each path.

The same construction as the one above works also for spin. Although the details are outside the scope of the present paper, we sketch the idea for spin ½: Here the paths are "rotational" paths in $\mathbb{R}^3 \times SO(3)$, and the action is a sum at each instant along a rotational path associated with a classical "top" action and a magnetic-field action [38]. There are two homotopy classes of paths in $SO(3)$. When entangled spin ½ particles enter Stern-Gerlach (SG) devices on each side, the correspondingly congruent spin paths, apart from the measurements, coordinate the correlations, as with the congruent paths in the two-particle interferometer. The SP explanation of the quantum observables for spin ½ (postulate 6, Appendix A) has certain aspects in common with the explanation above for effects at a beamsplitter for spinless particles. This is discussed further in postulate 6. For other cases of spin, such as spin 1, there is a projection onto additional subclasses over the two homotopy classes.

## APPENDIX A: POSTULATES FOR A "HIDDEN REALITY" BASED ON ALL POSSIBLE PATHS IN A QUANTUM SYSTEM

Although the following postulates provide—in terms of all possible paths—a "hidden mechanism" underlying quantum phenomena, they are obviously consistent with experiment. They form the basis for a system that replicates the predictions of quantum mechanics, and Bell's theorem cannot rule them out on the grounds that they involve local realism, since they are for a nonmeasurable space.

**Postulate 1 (Shadow stream):** Whenever a tangible (ordinary) particle travels between two points it generates what we call a "shadow stream" of infinitely many counterpart shadow particles also traveling between the points, the tangible randomly taking one of the paths (we include the tangible particle in its associated shadow stream). As noted earlier, tangible particles only interact with shadow counterparts of the same type—tangible photons with shadow photons, tangible electrons with shadow electrons, etc. With each path we associate a unit vector in the complex plane, $e^{iS[\text{path}]/\hbar}$ —an "exponentiated-action" term.

**Postulate 2 (Amplitude as the sum of unit vectors):** The amplitude for a shadow stream generated by a tangible particle traveling between two points $x_1$ to $x_2$ is the sum of exponentiated-action terms over all possible paths between the points.

$$\sum_{\substack{\text{all paths } x(t) \\ \text{from } x_1 \text{ to } x_2}} \exp(iS[x(t)]/\hbar)$$

The exponentiated action acts like a "clock-pointer" that begins "ticking" at a regular rate as the particle proceeds along the path. Because the value of $\hbar$ is quite small, the ticking is at a very rapid rate. Here $\hbar$ is Planck's constant $h$ divided by $2\pi$, "path" is the path of the particle, and $S[\text{path}]$ is called the "action over the path." The action functional $S$ in coordinate space is based on the Lagrangian (integral over the difference of kinetic and potential energy). The action-functional approach has been known since the eighteenth century from the work of Euler and Lagrange to be equivalent to Newtonian mechanics in representing the motion of systems with algebraically described constraints [39].

The amplitude can be represented geometrically by the "long" vector in a Cornu spiral, as in Fig. 4 [40]. Vectors associated with the paths that take longer times for a particle to travel from the source to point $A$ in the figure tend to curl up in the spiral, canceling each other (pointing in opposite directions). On the other hand, vectors over paths of similar lengths tend to add up constructively in the vector sum—represented by the long vector in the Cornu spiral. The absolute square of the magnitude of the long vector is proportional to the probability that the tangible particle in the stream pursues the so-called path of least time.

**Postulate 3** For *two non-interacting tangible particles* the composite amplitude of the two streams generated by the tangibles is *the product of the separate amplitudes*. (We are stating this postulate for convenience; it actually follows by an application of the rule, "The amplitude that one particle will do one thing and the other one do something else is the product of the two amplitudes that the two particles would do the two things separately." [41])

Now we will state two postulates that deal with 50-50 beamsplitters. The beamsplitter postulates are designed to duplicate the bivalent functions that determine which path a classical object pursues when it encounters a fork along a path in the toy model of Sec. II. These postulates are intuitively obvious and consistent with the experimental outcomes of a Rarity-Tapster interferometer [42], although they are based on local realism.

**Postulate 4 (Beamsplitter):** For simplicity, we confine the discussion to $\mathbb{R}^2$ and consider beamsplitters in just one- or two-particle interferometers. With each beamsplitter we associate a rotation $\Gamma$ of the unit circle in $\mathbb{R}^2$ (an unknown variable, of course).

(a) **Single-particle interferometer**: Note that there is no concern regarding locality in such an interferometer, and the underlying space is in fact measurable. In a typical setup of this kind we have two 50-50 beamsplitters. The tangible particle travels first from a point source to a beamsplitter [43], and from there it has a choice of paths to a second beamsplitter. For $i = 1$, 2, let $\Gamma_i$ be the associated rotation for the $i^{\text{th}}$ beamsplitter. If the tangible's exponentiated-action is of the form $r_1 e^{i\theta_1}$ when it encounters the first beam splitter, then it is transmitted if $\theta_1$ is in $\Gamma_1(\{(\cos\phi,\sin\phi):0\leq\phi\leq\pi\})$, otherwise it is reflected. When the tangible reaches the second beamsplitter, let $r_2 e^{i\theta_2}$ be the sum over the two paths leading to the second beamsplitter, one path traveled by the tangible, the other by its shadow counterpart. If $\theta_2$ is in $\Gamma_2(\{(\cos\phi,\sin\phi):0\leq\phi\leq\pi\})$, otherwise it is reflected.

(b) **Two-particle interferometer**: Here we have infinitely many paths, and the underlying space is non-measurable. There is a necessary positional uncertainty in the source [44]; and, as noted in Sec. II, this leads to a product of amplitudes on each side, each separate amplitude being a sum over all possible paths in the related shadow stream. As in the single-particle case, the amplitude on a side is of the form $re^{i\theta}$. If the angle $\theta$ is in the interval $\Gamma(\{(\cos\phi,\sin\phi):0\leq\phi\leq\pi\})$, the tangible is transmitted, otherwise it is reflected.

**Postulate 5 (Twin pairs)**: The emission of two entangled tangible particles traveling in opposite directions generates two oppositely directed shadow streams (oppositely directed by conservation of momentum). These streams typically contain uncountably many particles accompanying the two twin daughter-tangibles. Moreover, for each particle (tangible or shadow) in one of the paired streams, there is a twin counterpart in the other stream whose associated unit vector at the source event is pointing in the same direction as that of its twin. These are the identical, initial "instructions" carried by the twin particles, different pairs possibly pointing in different directions at the start. As the twin particles travel to oppositely directed destination points, they follow "action instructions" (described further in Sec. II above).

**Postulate 6 (spin-½)** Consider a spin-½ tangible particle, such as an electron, moving with its shadow counterparts through the inhomogeneous magnetic field of a Stern-Gerlach (SG) device. In SP we view each particle in terms of local realism, picturing each particle as a tiny, spinning, top-like, dipole magnet. The tangible's shadow stream consists of infinitely many such tops spinning in various arbitrary directions, a continuum of possibilities, each top having corresponding simultaneous spin components in various directions.

In Fig. 5 below, we represent the tangible as the "darker" particle traveling in its shadow stream of infinitely many, paler, counterpart particles.

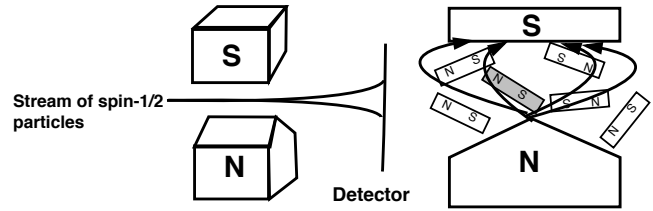

Fig. 5 The darker particle is a tangible particle traveling in a shadow homotopy class with other spin 1/2 particles (tiny dipole magnets)

The observables are the outcome of a sum of interference effects over all the paths associated with particles in the shadow stream, the tangible spinning in one of its two possible homotopy classes in the stream, where a spin path is in $\mathbb{R}^3 \times SO(3)$ ($SO(3)$ has just two homotopy classes of paths). The effects are quantified by the sum of exponentiated-action terms over the paths, where the action is a sum at each instant along a rotational path associated with a classical "top" action and a magnetic-field action [45].

There are just a small number of fundamental experimental results that must be accounted for by any mathematical representation of spin-½ [46, 47]. In standard quantum mechanics (SQM) this is done elegantly by a two-dimensional vector space over the complex numbers.

It is also done elegantly in SP by the path integral, viewing spin in terms of a shadow stream: When a spin-½ tangible particle travels within an SG field for the first time, it will travel randomly in one of two homotopy classes (for spin other than spin-½, one takes further projections onto the shadow stream). The sum of the exponentiated-action terms of all the particles in its homotopy class leads either to $+\hbar/2$ or $-\hbar/2$ on the SG detection plate. If the tangible exits a prior SG field, after traveling in a particular homotopy class, and then enters a field of the same orientation it will continue in that homotopy class to produce the same outcome as previously. But if it enters a field oriented in a different direction it will go randomly into one of two homotopy classes for the new orientation, producing the usual result. If the tangible exits an SG device and then enters a "modified-SG $\alpha$ device" (as defined by Feynman [48]) it will retain its previous orientation on exiting, because the sum of "exponentiated-$\alpha$" terms in the two homotopy classes of the stream of the modified setup will recombine to the entering value, leaving the particle oriented in its previous direction. In this way we see that the SP outcomes replicate the fundamental quantum mechanical predictions. This is similar to how we accounted for the effects at a beamsplitter above. Such observables are precisely represented in the two-dimensional Hilbert-space linear algebra of SQM by the two eigenvalues of self-adjoint operators on the system. As we have been emphasizing, however, the SQM representation is completely opaque as to the inner mechanism.

## APPENDIX B: SPACE OF ALL PATHS IS NOT MEASURABLE

As is well known, the space of all continuous paths joining two points in coordinate space is not measurable. For convenience, here is a proof (from [49]) that there is no nontrivial, translation invariant measure $\mu$ on the space $[a, b]^{[0, 1]}$, where $[a, b]^{[0, 1]}$ denotes the set of all continuous functions $x(t)$ from $[0, 1]$ onto $[a, b] \subseteq \mathbb{R}^d$, where $x(0) = a$ and $x(1) = b$.

First observe that in the space $[a, b]^{[0, 1]}$ every $r$-ball $B_r(y)$ (the open ball centered at the path $y(t)$ and having radius $r$) has the same measure as any other $r$-ball $B_r(z)$. Because let $y(t)$ be a path in the space. We are assuming that $\mu$ is translation invariant. Then the $r$-ball $B_r(y)$ is the translate of another $r$-ball $B_r(z)$ by $B_r(y) = B_r(z) + (y - z)$. Here $B_r(y)$ is the ball centered at the path $y$ in $[a, b]^{[0, 1]}$ with radius $r$.

Now let $O$ be a non-empty open set of finite $\mu$-measure in $[a, b]^{[0, 1]}$. Thus given any $x$ in $O$, there exists $r_\# > 0$ such that $B_{r_\#}(x) \subseteq O$. Then by the monotonicity property of a measure, $\mu(B_{r_\#}(x)) \leq \mu(O) < \infty$. By translation invariance of $\mu$, and again by monotonicity, we have for all $y$ in $[a, b]^{[0, 1]}$, $\mu(B_r(y)) < \infty$ whenever $0 < r \leq r_\#$.

Let $\|x\| = \max(\|x(t)\| : 0 \leq t \leq 1)$, where $\|x(t)\|$ is the Euclidean norm of $x(t)$ in $\mathbb{R}^d$ (the distance from the origin to $x(t)$). Then we define a sequence of functions $x_n$ in $[a, b]^{[0, 1]}$ as follows:

$$x_n(t) = \begin{cases} 0 & \text{if } 0 \leq t \leq 2^{-n} \\ 2^{n-1} r_\# (t - 2^{-n}) & \text{if } 2^{-n} \leq t \leq 2^{-n+1} \\ r_\# / 2 & \text{if } 2^{-n+1} \leq t \leq 1 \end{cases}$$

Thus $\|x_n\| = r_\# / 2$ and $\|x_n - x_m\| = r_\# / 2$ for $n \neq m$. Therefore $B_{r_\#/4}(x_n) \cap B_{r_\#/4}(x_m) = \emptyset$ if $n \neq m$ and $B_{r_\#/4}(x_n) \subseteq B_{r_\#}(0)$ for every $n$. Hence $B_{r_\#}(0)$ contains an infinite disjoint sequence of sets $\{B_{r_\#/4}(x_n)\}$ all having the same measure and such that $0 \leq \mu(B_{r_\#/4}(x_n)) < \infty$. Since $\mu(B_{r_\#}(0)) < \infty$, we have $\mu(B_{r_\#/4}(x_n)) = 0$ for every $n$. But $[a, b]^{[0, 1]}$ is separable. So there exists a countable dense set $\{y_j\}_{j=1}^{\infty}$. Thus $[a, b]^{[0, 1]} = \bigcup_{j=1}^{\infty} B_{r_\#/4}(y_j)$. Hence

$$\mu([a,b]^{[0,1]}) \leq \sum_{j=1}^{\infty} \mu(B_{r_\#/4}(y_j)) = 0.$$

Thus $\mu$ is necessarily identically 0, a trivial measure.

### APPENDIX C

In computing the probability for the correlations when $\alpha \neq \beta$, we suppose, say, $\alpha > \beta$, where of course by design each measurement is less than $2\pi$. Then, after the measurements, the objects arrive at the fork points with the clock-pointer on the right at some angle $\gamma$ while on the left it's at $\gamma + (\alpha - \beta)$, where $\alpha - \beta = 2\pi/3$ or $4\pi/3$. The probability that the objects end up on corresponding branches is (because of the bivalent functions $A$ and $B$) the probability that both pointers are between 0 and $\pi$, or both are between $\pi$ and $2\pi$. This probability in both cases is 0 when $\alpha - \beta = 4\pi/3$. When $\alpha - \beta = 2\pi/3$ we have, as noted, two possibilities: that both pointers are between 0 and $\pi$ or both are between $\pi$ and $2\pi$. Observe first that

$$\Pr(0 < \gamma < \pi/3) = (\pi/3)/2\pi = 1/6.$$

The product that both pointers are between 0 and $\pi$ is the product:

$$\Pr(0 < \gamma + 2\pi/3 < \pi) \cdot \Pr(0 < \gamma < \pi) = \Pr(0 < \gamma < \pi/3) \cdot \Pr(0 < \gamma < \pi) = (1/6)(1/2).$$

The same holds for the case where $\pi < \gamma < 2\pi$, and thus the probability is 1/6 when $\alpha > \beta$. Similarly, when $\beta > \alpha$. Therefore when $\alpha \neq \beta$ the total probability is 1/3.

[1] J. S. Bell, Physics I, 195-200 (1964) (reprinted in J. S. Bell, *Speakable and Unspeakable in Quantum Mechanics*, 2nd ed., Cambridge University Press, 2004 )
[2] M. Hoban, J. Wallman, and D. Browne, "Generalized Bell Inequality Experiments and Computation," Phys. Rev. A **84**, 062107 (2011), Dec. 6
[3] Y. Zhang, S. Glancy, and E. Knill, "Asymptotically optimal data analysis for rejecting local realism," Phys. Rev. A (accepted Nov 21, 2011)
[4] A. Acín, S. Massar, S. Pironio, Phys. Rev. Lett. **108**, 100402 (2012) – Published March 9, 2012
[5] J. S. Bell, "Bertlmann's socks and the nature of reality," *Speakable and Unspeakable in Quantum Mechanics*, 2nd ed., Cambridge University Press, 2004
[6] E. Pomarico, J. Bancal, B. Sanguinetti, A, Rochdi, N, Gisin, Phys. Rev. A **683**, 052104 (2011)
[7] N. Gisin, V. Scarani, A. Stefanov, A. Suarez, W. Tittel and H. Zbinden, Optics & Photonics News, 51, December 2002
[8] W. C. Myrvold, Philosophy of Science 70 (December 2003)
[9] S. Goldstein et al. (2011), Scholarpedia, 6(10):8378.
[10] T. Maudlin, Am. J. Phys. **78** (3), January 2010
[11] G. W. Johnson and M. L. Lapidus, *The Feynman Integral and Feynman's Operational Calculus*, Oxford Mathematical Monographs, Oxford University Press (2002), p. 32

[12] D. Deutsch, *The Fabric of Reality*, Penguin Press, 1998

[13] R.P. Feynman, "Space-Time Approach to Non-Relativistic Quantum Mechanics," Rev. Mod. Phys. 20, 367-387 (1948) reprinted in *Feynman's Thesis, A New Approach to Quantum Mechanics*, edited by Laurie M. Brown, World Scientific Publishing, 2005.

[14] L. Schulman, Phys. Rev. 176, 1558-1569 (1968)

[15] H. Kleinert, *Path Integrals in Quantum Mechanics, Statistics, Polymer Physics, and Financial Markets*, 2009, World Scientific 5th edition

[16] M. Gardner, J. A. Wheeler, *Quantum Theory and Quack Theory*, New York Review of Books, Vol. 26, Number 8, May 17, 1979

[17]

[18] L. Vaidman, "Many Worlds Interpretation of Quantum Mechanics," Stanford Encyclopedia of Philosophy, 2002

[19] T. Maudlin, Am. J. Phys. **78** (3), January 2010

[20] A. Einstein, B. Podolsky, N. Rosen, Phys. Rev. 47, 777 (1935).

[21] D.M. Greenberger, M. A. Horne, A. Shimony, A. Zeilinger, Am. J. Phys. 58 (12), December 1990

[22] J. S. Townsend, *A Modern Approach to Quantum Mechanics*, Ch 1, University Science Books, 2000; or R. P. Feynman, R. B. Leighton, and M. Sands, *The Feynman Lectures on Physics* (Addison-Wesley, Reading, MA, 1964), Vol. III, Ch 5

[23] J. S. Bell, *Speakable and Unspeakable in Quantum Mechanics*, 2nd ed., Cambridge University Press, 2004

[24] N. David Mermin, Shedding (red and green) light on "time related hidden parameters", e-print quant-ph/ arXiv:0206118v1 18 Jun 2002

[25] D. Griffiths, *Introduction to Quantum Mechanics*, 2$^{nd}$ edition, p. 13, Pearson Prentice Hall

[26] H. J. Bernstein, D. M. Greenberger, M. A. Horne, A. Zeilinger, "Bell theorem without inequalities for two spinless particles," Physical Review A, Vol. 47, Number 1, January, 1993

[27] J. S. Bell, Physics I, 195-200 (1964) No doubt resigned to what he took to be the unbreakable nonlocality consequences of his theorem, Bell was a proponent of Bohm's theory, although describing it as a "grossly nonlocal theory."

[28] N. D. Mermin, *Bringing home the atomic world: Quantum mysteries for anybody*, Am. J. Phys. **49**(10),Oct. 1981

[29] Based on the diagram in H. J. Bernstein, D. M. Greenberger, M. A. Horne, A. Zeilinger, "Bell theorem without inequalities for two spinless particles," Physical Review A, Vol. 47, Number 1, January, 1993

[30] R. P. Feynman, R. B. Leighton, and M. Sands, *The Feynman Lectures on Physics* (Addison-Wesley, Reading, MA, 1964), Vol. III, Ch 1.

[31] D. M. Greenberger, M. A. Horne, A. Zeilinger, Physics Today, August 1993

[32] D. M. Greenberger, M. A. Horne, A. Zeilinger, Physics Today, August 1993

[33] M. O. Scully, K. Druhl, Phys. Rev. A 25, 2208-2213 (1982)

[34] V. Scarani, A. Suarez, Am. J. Phys. **66** (8), August 1998

[35] R. P. Feynman, QED: *The Strange Theory of Light and Matter*, Princeton University Press, Princeton, NJ, 1986.

[36] Based on a similar figure in R. P. Feynman, QED: *The Strange Theory of Light and Matter*, Princeton University Press, Princeton, NJ, 1986. An excellent elementary discussion of Feynman's approach is also in J. Ogborn and E. F. Taylor "Quantum physics explains Newton's laws of motion," Physics Education, January, 2005

[37] J. Bernstein, *Quantum Profiles* (Princeton University Press), 1991, p. 82

[38] A. Atland, B. Simmons, *Condensed Matter Field Theory*, Cambridge University Press, 2006, Sec. 3.3.5

[39] D. Morin, *Introduction to Classical Mechanics: With Problems and Solutions, Cambridge University Press*, 2007, Ch. 6

[40] Based on R. P. Feynman, QED: *The Strange Theory of Light and Matter*, Princeton University Press, Princeton, NJ, 1986. An excellent elementary discussion of Feynman's approach is also in J. Ogborn and E. F. Taylor "Quantum physics explains Newton's laws of motion," Physics Education, January, 2005

[41] R. P. Feynman, R. B. Leighton, and M. Sands, *The Feynman Lectures on Physics* (Addison-Wesley, Reading, MA, 1964), Vol. III, Ch 3

[42] There is some evidence for this is (although it is not the intention of the paper's authors) in H. J. Bernstein, D. M. Greenberger, M. A. Horne, A. Zeilinger, "Bell theorem without inequalities for two spinless particles," Physical Review A, Vol. 47, Number 1, January, 1993

[43] V. Scarani, A. Suarez, Am. J. Phys. **66** (8), August 1998

[44] D. M. Greenberger, M. A. Horne, A. Zeilinger, Physics Today, August 1993

[45] A. Atland, B. Simmons, *Condensed Matter Field Theory*, Cambridge University Press, 2006, Sec. 3.3.5

[46] As described, for instance, in J. S. Townsend, *A Modern Approach to Quantum Mechanics*, Ch 1, University Science Books, 2000

[47] R. P. Feynman, R. B. Leighton, and M. Sands, *The Feynman Lectures on Physics* (Addison-Wesley, Reading, MA, 1964), Vol. III, Ch 5

[48] R. P. Feynman, R. B. Leighton, and M. Sands, *The Feynman Lectures on Physics* (Addison-Wesley, Reading, MA, 1964), Vol. III, Ch 5

[49] G. W. Johnson and M. L. Lapidus, *The Feynman Integral and Feynman's Operational Calculus*, Oxford Mathematical Monographs, Oxford University Press (2002), p. 32